\documentclass[twocolumn]{aastex62}

\graphicspath{{./}{figures/}}

\received{-}
\revised{-}
\accepted{-}
\submitjournal{ApJ}

\shorttitle{APEX observations of V883 Ori}
\shortauthors{White et al.}

\begin{document}

\title{APEX Observations of the CO Envelope around the Young FUor-type Star V883 Ori}

\correspondingauthor{Jacob Aaron White}
\email{jacob.white@csfk.mta.hu}

\author[0000-0001-8445-0444]{J. A. White}
  \affiliation{Konkoly Observatory, Research Centre for Astronomy and Earth Sciences, Hungarian Academy of Sciences, Konkoly-Thege Mikl\'os \'ut 15-17, 1121 Budapest, Hungary}

\author[0000-0001-7157-6275]{\'A. K\'osp\'al}
  \affiliation{Konkoly Observatory, Research Centre for Astronomy and Earth Sciences, Hungarian Academy of Sciences, Konkoly-Thege Mikl\'os \'ut 15-17, 1121 Budapest, Hungary}
  \affiliation{Max Planck Institute for Astronomy, K\"onigstuhl 17, 69117 Heidelberg, Germany}

\author[0000-0003-1817-6576]{C. Rab}
  \affiliation{Kapteyn Astronomical Institute, University of Groningen, Postbus 800, 9700 AV Groningen, The Netherlands}

\author[0000-0001-6015-646X]{P. \'Abrah\'am}
  \affiliation{Konkoly Observatory, Research Centre for Astronomy and Earth Sciences, Hungarian Academy of Sciences, Konkoly-Thege Mikl\'os \'ut 15-17, 1121 Budapest, Hungary}

\author[0000-0002-4283-2185]{F. Cruz-S\'aenz de Miera}
  \affiliation{Konkoly Observatory, Research Centre for Astronomy and Earth Sciences, Hungarian Academy of Sciences, Konkoly-Thege Mikl\'os \'ut 15-17, 1121 Budapest, Hungary}

\author[0000-0002-6018-1371]{T. Csengeri}
  \affiliation{Max-Planck-Institut f\"ur Radioastronomie, Auf dem H\"ugel 69, 53121 Bonn, Germany}

\author[0000-0003-3453-4775]{O. Feh\'er}
  \affiliation{Konkoly Observatory, Research Centre for Astronomy and Earth Sciences, Hungarian Academy of Sciences, Konkoly-Thege Mikl\'os \'ut 15-17, 1121 Budapest, Hungary}
  \affiliation{IRAM, 300 Rue de la Piscine, 38046 Saint Martin d'H\`eres, France}

\author{R. G\"usten}
 \affiliation{Max-Planck-Institut f\"ur Radioastronomie, Auf dem H\"ugel 69, 53121 Bonn, Germany}

\author[0000-0002-1493-300X]{T. Henning}
  \affiliation{Max Planck Institute for Astronomy, K\"onigstuhl 17, 69117 Heidelberg, Germany}

\author{E. Vorobyov}
  \affiliation{University of Vienna, Dept. of Astrophysics, T\"urkenschanzstr. 17, 1180 Wien, Austria}
  \affiliation{Research Institute of Physics, Southern Federal University, Stachki 194,
Rostov-on-Don, 344090, Russia}

\author{M. Audard}
  \affiliation{Department of Astronomy, University of Geneva, Ch. d'Ecogia 16, 1290 Versoix, Switzerland}

\author{A. Postel}
  \affiliation{Department of Astronomy, University of Geneva, Ch. d'Ecogia 16, 1290 Versoix, Switzerland}



\begin{abstract}

The accretion-driven outbursts of young FU Orionis-type stars may be a common stage of pre-main sequence evolution and can have a significant impact on the circumstellar environment as it pertains to the growth of solids and eventually planets. This episodic accretion is thought to be sustained by additional gas in-falling from the circumstellar envelope and disk. We present APEX observations of the CO gas in the envelope around V883 Orionis, a young outbursting star. The observations mapped the $^{12}$CO(4--3), $^{12}$CO(3--2), and $^{13}$CO(3--2) lines with the FLASH$^{+}$ instrument and the $^{12}$CO(6--5) line with the SEPIA instrument. We detected high signal-to-noise emission extending out to radii $>10000$ au and calculated integrated fluxes of $1100~\rm Jy~km~s^{-1}$ for $^{12}$CO(6--5), $2400~\rm Jy~km~s^{-1}$ for $^{12}$CO(4--3), $1600~\rm Jy~km~s^{-1}$ for $^{12}$CO(3--2), and $450~\rm Jy~km~s^{-1}$ for $^{13}$CO(3--2). We used the thermo-chemical code P{\small RO}D{\small I}M{\small O} to test several models and find the data are best described by an envelope structure with  $\rm M_{env}\approx 0.2-0.4\,M_{\odot}$ and a mass-infall rate of $\rm \dot{M}_{inf}=1-2\times10^{-6}\,M_{\odot}\,yr^{-1}$. We infer that the observed envelope and outflow structure around V883 Ori could be caused by multiple outbursts, consistent with episodic accretion.

\end{abstract}



\section{Introduction} \label{sec:intro}

During a star's pre-main sequence evolution, material will be accreted from the disk onto the protostar, fueling its growth. A common stage of this evolution may be periods of significantly enhanced mass infall, known as episodic accretion. This is evidenced by outbursts that increase a system's apparent brightness by several magnitudes and total luminosity by as much as $10-100\times$. Understanding the underlying mechanisms that cause the observed outbursts of young stellar objects (YSOs) is important not only for building a complete picture of star formation, but also for the potential implications on the planet formation process. Young outbursting systems can be categorized as either FUors or EXors. FUors, named after FU Orionis \citep{herbig66}, are embedded systems that have very strong outbursts that can last $>100$ yr. These stars will have optical spectra similar to rapidly rotating earlier-type giant stars and infrared (IR) spectra more similar to M-giant stars \citep{hartmann96}. EXors, named after EX Lupi \citep{herbig89}, have outbursts that are weaker and shorter lived as compared to FUors. Indeed, FUors and EXors could represent different evolutionary stages of the same types of objects as EXors typically have lower disk masses and are no longer embedded. The likely cause of both these types of outbursts is episodic accretion of material from the circumstellar disks onto the protostars. The transport mechanism(s) that delivers additional material onto protostars, however, is still unknown \citep[see][for a review of outbursting YSOs]{audard14}.

Winds and molecular outflows are common components of star formation and disk evolution \citep{williams11}. These components can carry away angular momentum, allowing material in the circumstellar disk to accrete onto the star.  As material is transported inward, the disk is thought to be replenished by material from the large envelopes, which can extend to 1000s of au \citep[e.g.,][]{cassen81, terebey84}. This ``refueling" of material can support the episodic nature of the accretion events. Therefore, studying the envelopes of FUors is crucial to understanding the nature of accretion events in young stars \citep[e.g.,][]{quillen05}. The envelopes of FUors can be inferred through absorption seen in the $10~\mu$m silicate feature \citep{quanz07} or through IR excess \citep[e.g.,][]{green06}. Direct observations of the envelope gas, however, are less common but crucial to constraining the mass and morphological structure \citep{Kospal2017}. An accurate characterization of the envelope around an FUor can determine the timescale and strength of the mass infall rate, will yield insight into how many future outburst events can be sustained, and can help distinguish the difference between the circumstellar environments of FUors and that of other YSOs.

V883 Ori is an embedded $1.3\pm0.1~\rm M_{\odot}$ \citep{cieza16} pre-main sequence star located in the Lynds 1641 cloud in the Orion star forming region at a distance of $414\pm7$ pc \citep{menten07}. \citet{strom93} first suggested that V883 Ori may be an FUor as it shares similar optical spectral features to FU Orionis and has a flat  spectral energy distribution (SED). High resolution near-IR spectra further indicate that V883 Ori is best classified as an FUor \citep{greene08}. While V883 Ori has not been observed in a pre-outburst state, analysis of photographic plates indicate the outburst likely began before 1888 \citep{pickering1890}. Coupled with a potential ALMA-resolved water snowline \citep{cieza16} and the presence of complex organic molecules \citep{vant18, lee19}, V883 Ori is a key testbed system for understanding outbursts and the early stages of star formation and disk evolution. 

In this paper, we present APEX FLASH$^{+}$ and SEPIA B9 observations of the CO in V883 Ori's envelope. In Sec.\,\ref{sec:obs}, we describe the data reduction and observational results. In Sec.\,\ref{sec:env}, we detail the envelope model. In Sec.\,\ref{sec:discussion}, we discuss the implications of the model. In Sec.\,\ref{sec:summary}, we summarize the results.

\begin{table*}
\begin{center}
\caption{Summary of our APEX observations. The line $\sigma_{rms}$ is from the line-free channels, re-binned to $1.0~\rm km~s^{-1}$. The peak, FWHM, and envelope flux (Env. Flux) of a given line is taken from fitting a Gaussian to the lower-velocity peak in the integrated spectra in Fig.\,\ref{fig:spec}. The uncertainties on the peak and FWHM are the $1\sigma$ uncertainties from the co-variance matrix in the Gaussian fitting. The uncertainties on the envelope flux are $\sigma_{rms}$, model fitting uncertainty, and absolute flux calibration uncertainty added in quadrature. The absolute flux calibration uncertainty is taken as $15\%$ for $^{12}$CO(6-5) and $10\%$ for all other lines and is the dominate source of flux uncertainty. The beam efficiencies and $\rm Jy~K^{-1}$ conversion factors for each observational setup are given in Sec.\,\ref{sec:obs}.}\label{obs_overview}
\bigskip
\begin{tabular}{c|c c c c c c c c}
\hline
  Species & Observation Date & Frequency &  Beam & $\sigma_{rms}$ & Ch. Width & Peak & FWHM & Env. Flux\\
   & YYYY MMM DD & [GHz] & [$''$] & [Jy beam$^{-1}$] & [$\rm km~s^{-1}$] & [Jy beam$^{-1}$] & [$\rm km~s^{-1}$] & [Jy km s$^{-1}$]\\
\hline\hline
 $^{12}$CO(6--5)   & 2018 OCT 15/21 & 691.473 & 9.6 & 2.2 & 0.20  & $1100\pm45$ & $0.38\pm0.02$ & $1100\pm170$ \\
 $^{12}$CO(4--3)   & 2016 SEP 06 & 461.041 & 14.4 & 1.4 & 0.050 & $2800\pm20$ & $0.35\pm0.01$ & $2400\pm240$ \\
 $^{12}$CO(3--2)   & 2016 SEP 06 & 345.796 & 19.2 & 0.53 & 0.033 & $1800\pm10$ & $0.36\pm0.01$ & $1600\pm160$\\
 $^{13}$CO(3--2)   & 2016 SEP 06 & 330.588 & 20.1 & 0.81 & 0.033 & $680\pm6$ & $0.26\pm0.01$ & $450\pm45$ \\
 
\end{tabular}
\end{center}
\end{table*}

\section{Observations}\label{sec:obs}
We observed V883 Ori with the Atacama Pathfinder EXperiment (APEX), a 12\,m diameter millimeter-wave telescope located on the Llano de Chajnantor in the Chilean Atacama desert \citep{gusten06}. We used the FLASH$^+$ receiver to measure the $^{12}$CO(3--2), $^{13}$CO(3--2), and $^{12}$CO(4--3) lines on 2016 September 6 (Project 098.F-9505(A), PI: Th.~Henning), and the SEPIA B9 receiver to measure the $^{12}$CO(6--5) line on 2018 October 15 and October 21 (Project 0102.F-9802, PI: J.A.~White). The SEPIA observations were part of the Science Verification of the B9 receiver. For each line, we used an On-The-Fly (OTF) mapping procedure centered at the V883 Ori J2000 coordinates of R.A. = 05$^{h}$38$^{m}$18$^{s}$.097 and DEC. = $-07^{\circ}02'25\farcs96$. The main parameters of our observations are summarized in Table\,\ref{obs_overview}.

FLASH$^+$ is a dual-frequency heterodyne receiver, operating simultaneously in the 345\,GHz and 460\,GHz atmospheric windows, providing 4\,GHz bandwidth in each sideband \citep{klein14}. The lower frequency channel was tuned to 344.2\,GHz in the upper sideband (USB) to simultaneously cover the $^{13}$CO(3--2) and $^{12}$CO(3--2) lines at 330.588\,GHz and 345.796\,GHz, respectively. The higher frequency channel was tuned to the $^{12}$CO(4--3) line at 461.041\,GHz in USB. We used the XFFTS backends, providing a 38\,kHz ($0.033~\rm km~s^{-1}$) spectral resolution for both the 3--2 lines and 76\,kHz ($0.050~\rm km~s^{-1}$) for the 4-–3 line. We obtained 120$''{\times}$120$''$ OTF maps using a relative reference off position 2500$''$ away in R.A. Before the observation, we checked in total power mode that the off position was clean of CO emission. The precipitable water vapor (PWV) was on average 0.5\,mm during our observations.

SEPIA is the Swedish-ESO PI receiver for APEX \citep{belitsky18}. The SEPIA Band 9 receiver is a dual polarization 2SB receiver \citep{hesper18,baryshev15} built to the specifications of the ALMA Band 9 receiver (600--720\,GHz). Currently, the XFFTS backend records 4\,GHz bandwidth in each sideband. The USB was centered on the $^{12}$CO(6--5) line at 691.473\,GHz with 450\,kHz ($0.20~\rm km~s^{-1}$) resolution. We obtained 150$''{\times}$150$''$ OTF maps, while the relative reference off position was 1000$''$ away in R.A. The PWV was 0.5--0.7\,mm during our two observations.

The data were reduced using the GILDAS CLASS package\footnote{http://www.iram.fr/IRAMFR/GILDAS/} and a first-order baseline was removed from the spectra. The main beam efficiencies determined by observing Mars are 0.73, 0.60, and 0.49 (APEX Team, private communication) at 352, 464, and 691\,GHz, respectively. The data were converted to Jy using 41\,Jy\,K$^{-1}$, 48\,Jy\,K$^{-1}$, and 79\,Jy\,K$^{-1}$ at 352, 464, and 691\,GHz, respectively. The telescope's beam size is 20$\farcs$1, 19$\farcs$2, 14$\farcs$4, and 9$\farcs$6 for the $^{13}$CO(3-–2), $^{12}$CO(3--2), $^{12}$CO(4--3), and $^{12}$CO(6--5) lines, respectively. The $\sigma_{rms}$ for each line is calculated by re-binning each spectrum to a $1~\rm km~s^{-1}$ channel width, then calculating the root-mean-square of all the line-free channels. The resulting values are summarized in Table\,\ref{obs_overview}. The absolute flux uncertainty is still unconstrained for these instruments but is taken as $10\%$ at 352 GHz and 464 GHz and as $15\%$ at 691 GHz. 

\begin{figure*}
    \centering
    \includegraphics[width=1.1\textwidth]{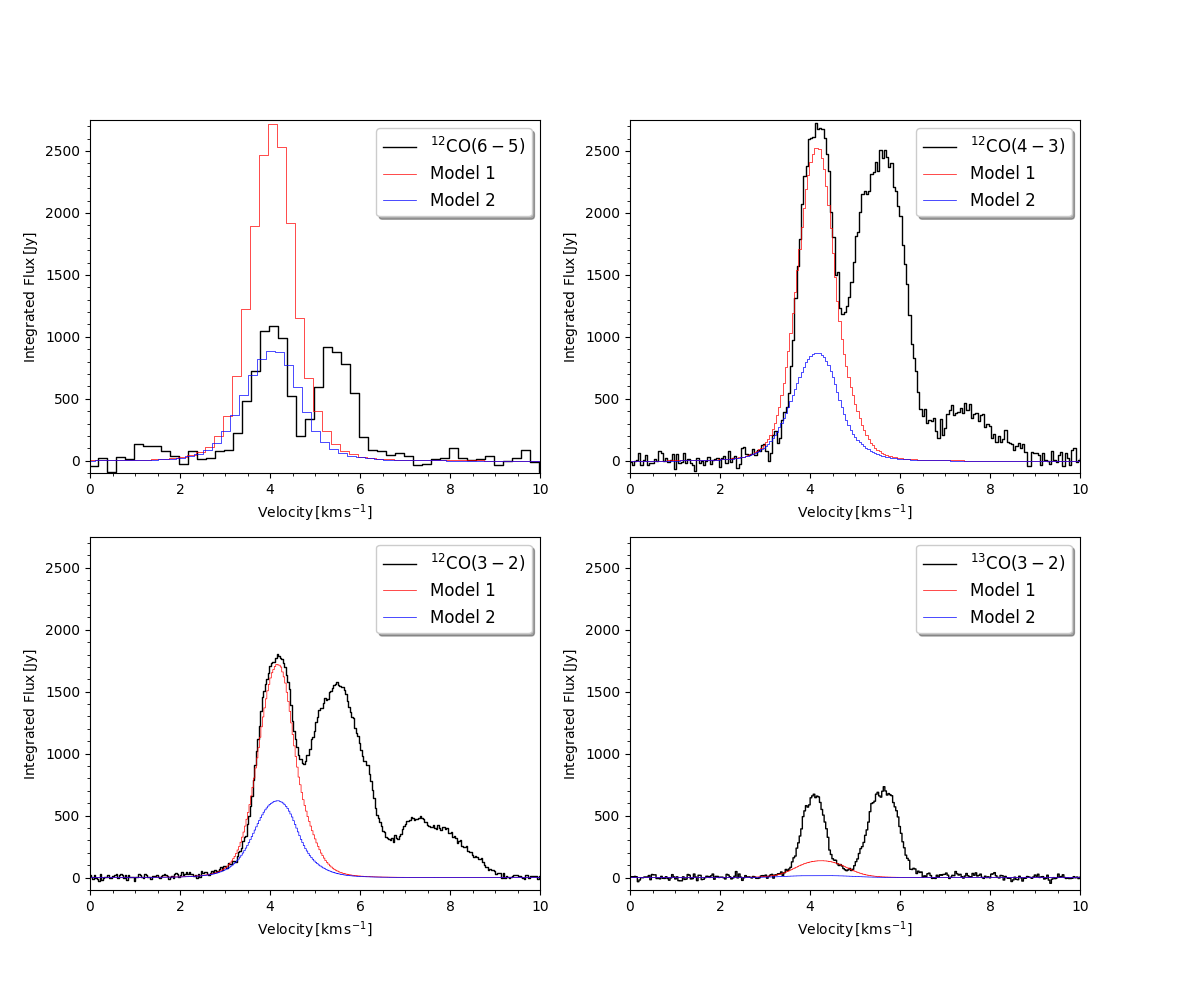}
    \caption{Integrated spectra of all the CO species over a 10000 au radius aperture. The black lines are the observed spectra. The red line is Model 1, with a mass-infall rate of $\dot{M}_\mathrm{inf}=2\times10^{-6}\,M_\mathrm{\odot}\,\mathrm{yr^{-1}}$. The blue line is Model 2, with $\dot{M}_\mathrm{inf} =1\times10^{-6} \,M_\mathrm{\odot} \,\mathrm{yr^{-1}}$. The lower-velocity peak for each species is from the envelope. The additional peaks seen for all species at $\rm \sim5.5~km~s^{-1}$ and for $^{12}$CO(4--3) and $^{12}$CO(3--2) at $\rm \sim7.5~km~s^{-1}$ are likely due to either outflow material or contamination from the surrounding molecular cloud. This is further illustrated in the velocity channel maps in the Appendix  (Figs.\,\ref{fig:vel_chan65}, \ref{fig:vel_chan43}, \ref{fig:vel_chan32}, and \ref{fig:vel_chan1332}). }
    \label{fig:spec}
\end{figure*}
\subsection{Observational Results}
The spectra of each molecular species is presented in Fig.\,\ref{fig:spec}, spatially integrated over a region of 10000 au. The spectra from V883 Ori's envelope are expected to have a single peak following an approximately Gaussian profile given the resolution of the observations. However, there are clearly 2 peaks in the emission seen in $^{12}$CO(6--5) and $^{13}$CO(3--2) and 3 peaks seen in $^{12}$CO(4--3) and $^{12}$CO(3--2). The best fit positions of the lower-velocity peak in each spectrum are all centered at $\sim4.1~\rm km~s^{-1}$ (the approximate systemic velocity). The other emission features are likely a combination of outflow material \citep[which flows along the SE/NW axis;][]{ruiz17a}, a non-spherical envelope structure, or diffuse emission from the surrounding molecular cloud. This is best seen in the velocity channel maps of each line (see Appendix \ref{sec:vel_app}). For example, Fig.\,\ref{fig:vel_chan65} shows the $^{12}$CO(6--5) line channel map. In this figure, the emission from the envelope (lower-velocity peak in the spectra) is largely contained within radii $<10000$ au at velocities $\rm <5.0~km~s^{-1}$ while the emission at velocities $\rm >5.0~km~s^{-1}$ is more consistent with diffuse emission at larger radial separations. The emission of the higher-velocity components fills the field of view in several velocity channels and does not show a concentration or a peak towards the position of V883 Ori. Those structures might be velocity coherent structures within the molecular cloud complex L 1641, similar to the ``fibers" in the Orion Integral Filament mapped in detail by \citet{Hacar2018}.

\begin{figure*}
    \centering
    \includegraphics[width=\textwidth]{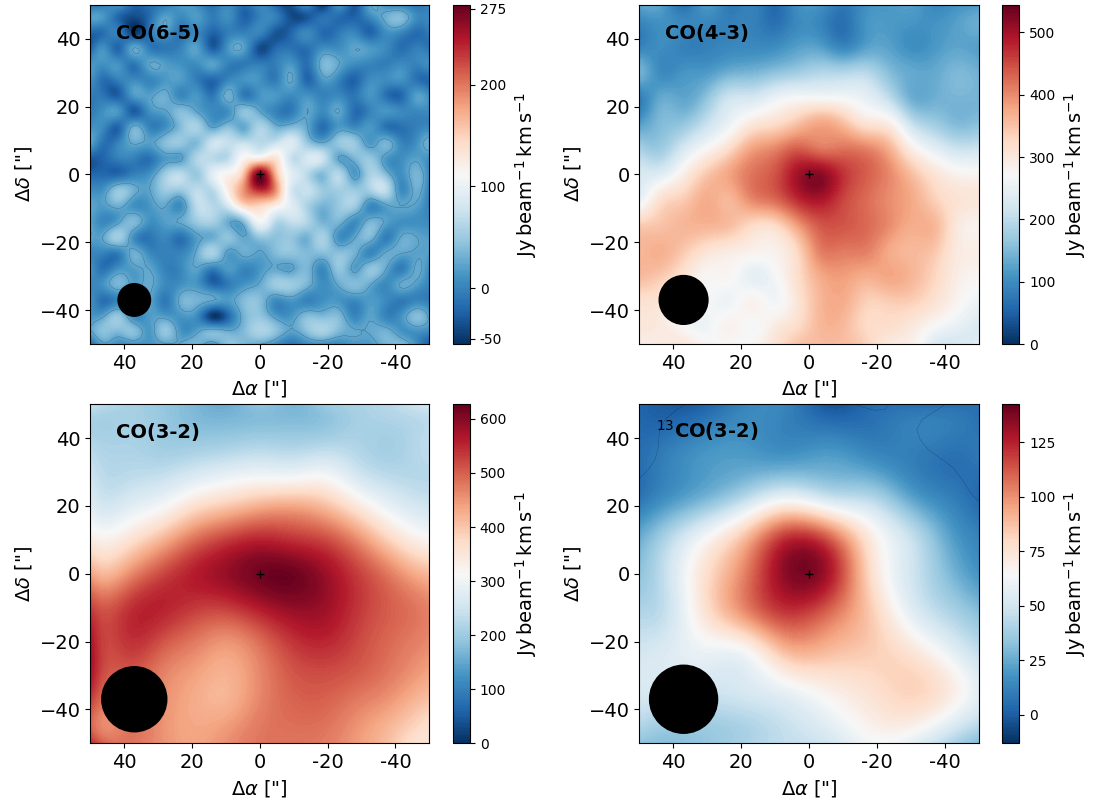}
    \caption{Integrated intensity maps for all of the observed CO lines. Each spectra was integrated from $\rm 2.4~km~s^{-1}$ to $\rm 4.8~km~s^{-1}$, which corresponds to the lower-velocity peak as seen in Fig.\,\ref{fig:spec}. The position of the star is denoted by the black plus and the beam size by the black circle. The $3\sigma$ contours are indicated by the black lines. We note that the contours are not visible for the $^{12}$CO(4--3) and $^{12}$CO(3--2) as the envelope emission is detected at a very high signal-to-noise.}
    \label{fig:int_map}
\end{figure*}

The compact emission in the channels near to the systemic velocity (e.g.,  $3.4~\mathrm{km~s^{-1}}$ and $4.6~\mathrm{km~s^{-1}}$ in Fig.\,\ref{fig:spec}) suggests in-falling material, as higher velocities are expected closer to the center (i.e., free-fall velocity $\propto 1/\sqrt r$). To calculate the peak, FWHM, and total envelope flux for each species, we first integrated each spectrum over a 10000 au radius aperture centered on V883 Ori. A multi-component Gaussian was fit to each spectrum in order to properly differentiate between emission from the envelope and the other peaks in the spectra. The best fit parameters from the lower-velocity Gaussian as seen in Fig.\,\ref{fig:spec} are summarized in Table\,\ref{obs_overview}. The integrated intensity maps ($0^{th}$ order moment) for each species are shown in Fig.\,\ref{fig:int_map}. Each of these maps only shows the integrated intensity from the lower-velocity peak of each spectra.

\section{Envelope Model}\label{sec:env}
We present models for the envelope structure applying the radiation thermo-chemical code P{\small RO}D{\small I}M{\small O} \citep{Woitke2009,Woitke2016}. This code self-consistently solves for the local radiation field, temperature, and chemical abundances for a given fixed axis-symmetric two-dimensional dust and gas density structure, and produces synthetic SEDs \citep{Thi2011} and spectral lines \citep{Woitke2011}. To model the density structure, we use the rotating in-falling envelope model of \citet{Ulrich1976} which has the mass of the protostar, $M_\mathrm{*}$, and the mass infall rate of the envelope, $\dot{M}_\mathrm{inf}$, as input parameters \citep[see][for further details]{Rab2017}. 

For V883 Ori, \citet{cieza16} derived $\rm M_*= 1.3\,M_{\odot}$ from ALMA $\rm C^{18}O(2-1)$ observations that resolve the inner Keplerian disk. These observations also constrain the disk's inclination to $38\degr$ and the systemic velocity to $4.3\,\mathrm{km\,s^{-1}}$. We do not model the disk structure as the APEX observations are not sensitive to the disk emission because of beam dilution. Our model also includes an empty outflow cavity with an opening angle of $75\degr$ and has an extension of $r=18 000\,\mathrm{au}$. 

\begin{figure}
    \includegraphics[width=0.48\textwidth]{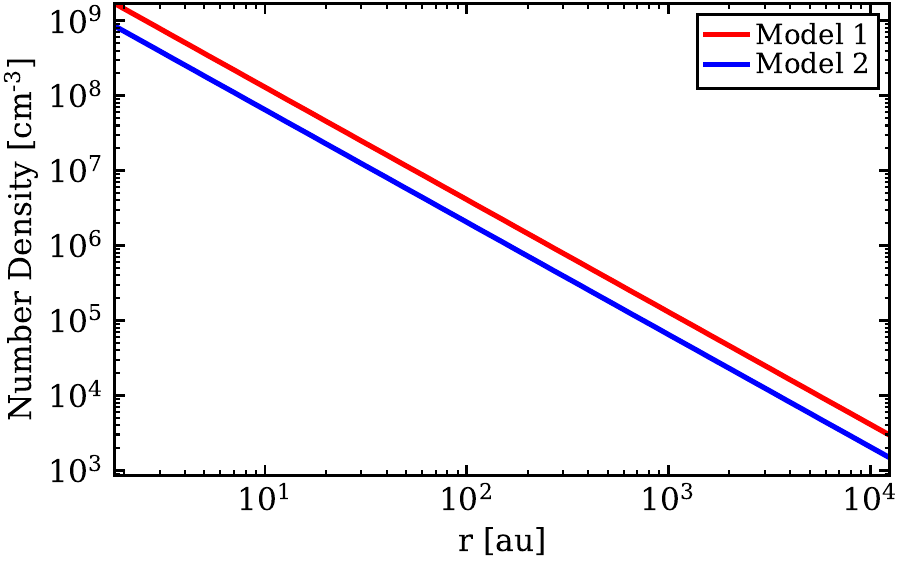}\hspace{0.3cm}
    \includegraphics[width=0.48\textwidth]{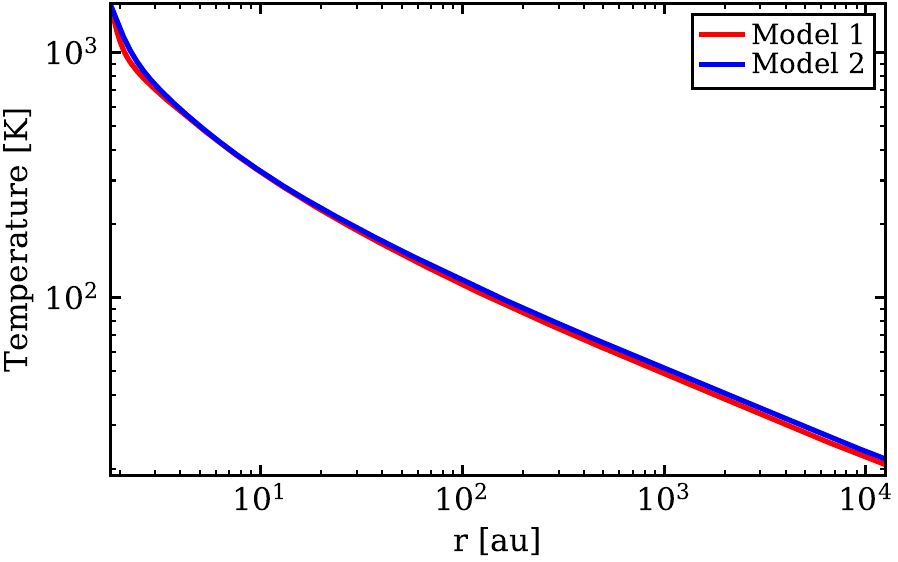}
    \caption{Radial number density (top panel) and temperature (bottom panel) profiles in the midplane of the two-dimensional envelope models. Model~1 is indicated by the red curve and Model 2 is indicated by the blue curve.}
    \label{fig:model}
\end{figure}

The outburst of V883 Ori is represented by a point-like source with $\rm L=200\,L_{\odot}$ in the center of the structure \citep[see also,][]{Rab2017}. The radial temperature and density profiles through the midplane of our models are shown in Fig.~\ref{fig:model}. These profiles are very similar to the Class I outburst models of \citet{Rab2017} and the 1D outburst models of \citet{Visser2015} for embedded targets. To compare our model to the data, we produced synthetic line cubes, convolved them with a Gaussian beam (size according to the data), and re-binned them to the velocity resolution of the observations. Further details on the modeling procedure and a comparison to the dust modelling of \citet{Furlan2016} are presented in Appendix~\ref{sec:modelcomparison}.

We note that our model does not cover to the entire area mapped by the observations, but does cover the region traced by $\mathrm{CO(6-5)}$ and $\mathrm{^{13}CO(3-2)}$ which show approximately symmetric emission in the lower-velocity component (see Fig.~\ref{fig:int_map}). $\mathrm{CO(4-3)}$ and $\mathrm{CO(3-2)}$ show also asymmetric structures especially towards the north-east (Fig.~\ref{fig:int_map}), which we cannot account for in our symmetric envelope model. Therefore, we only compare our model results within a radius of $10 000\,\mathrm{au}$. However, the observations might still pick up emission along the line of sight from structures that may or may not be part of V883\,Ori and are not included in our models. This could lead to our model overestimating the flux from the inner region.

Using our model framework, we can test if a simple envelope structure with the above constraints is consistent with the APEX data. In Fig.~\ref{fig:spec}, we show the results for two models with $\dot{M}_\mathrm{inf}=2\times10^{-6}\,\rm M_{\odot}\, yr^{-1}$ (Model~1) and  $\dot{M}_\mathrm{inf}=1\times10^{-6}\, \rm M_{\odot}\, yr^{-1}$ (Model~2). Model 1 is a good match for the lower-velocity peak of the $^{12}$CO(4--3) and $^{12}$CO(3--2) lines. The modeled velocity FWHM and peak are in good agreement with the data. This indicates that the lower-velocity peak of the profile is indeed consistent with an in-falling envelope positioned on top of the disk of V883 Ori. Our model also shows that a single envelope structure is not sufficient to explain all the peaks in the spectra. Assuming that self-absorption produces the double-peaked profile $\rm v_{sys}\approx5\,km\,s^{-1}$ is required, which is inconsistent with the value derived from ALMA disk observations. Furthermore, the FWHM of this single component ($\rm \sim 2\,km\,s^{-1}$) is too broad for a single in-falling envelope structure and the two components are not symmetric (e.g., the FWHM of the right component is a factor of 1.5 too wide). 

As seen in Fig.~\ref{fig:spec}, Model 1 underestimates the $^{13}$CO(3--2) by a factor of $\sim5$ and overestimates the $^{12}$CO(6--5) by a factor of $\sim2$. Model 2 on the other hand is a better match for $^{12}$CO(6--5), but underestimates all other lines. From the models, we can determine the main line emitting volume (i.e., the origin of 50\% of the emission) for each individual line. For all lines, this region is within the inner $10 000\,\mathrm{au}$. For these  emitting regions we can estimate averaged quantities such as the particle number density $n$, temperature $T$ and CO molecular abundances $\epsilon$. For Model 1, we find for the 
\mbox{$^{12}$CO(3--2)} line:~$n\!\approx\!8.7\!\times\!10^3\,\mathrm{cm^{-3}}$,  $\epsilon\!\approx\!2.5\!\times\!10^{-5}$, $T\!\approx\!26.0\,\mathrm{K}$; 
for \mbox{$^{13}$CO(3--2)}:~$n\!\approx2.7\!\times\!10^4\,\mathrm{cm^{-3}}$,   $\epsilon\!\approx\!1.1\!\times\!10^{-4}$, $T\!\approx\!33.3\,\mathrm{K}$; 
for \mbox{$^{12}$CO(4--3)}:~$n\!\approx\!1.3\!\times\!10^4\,\mathrm{cm^{-3}}$,  $\epsilon\!\approx\!4.6\!\times\!10^{-5}$, $T\!\approx\!28.5\,\mathrm{K}$ and 
for \mbox{$^{12}$CO(6--5)}:~$n\!\approx\!2.7\!\times\!10^4\,\mathrm{cm^{-3}}$,  $\epsilon\!\approx\!1.1\!\times\!10^{-4}$, $T\!\approx\!33.7\,\mathrm{K}$. 
This shows that in the model \mbox{$^{13}$CO(3--2)} and \mbox{$^{12}$CO(6--5)} trace similar regions that are warmer and denser compared to the other two lines. We explored varying other parameters in the model (e.g., central luminosity, dust opacities) but did not find a significantly better model that reproduces the data. This may be because our envelope density structure is too simple or our model does not correctly predict the temperature in the inner regions. All of the models tested under-predict the \mbox{$^{13}$CO(3--2)} significantly, which might indicate that  \mbox{$^{13}$CO(3--2)} does not only trace the envelope but also the SE/NW outflows reported by \citet{ruiz17a}. We discuss possible interpretations of this in Sect.~\ref{sec:discussion}.

The envelope masses in our models are approximately 0.4 and $0.2\,\rm M_{\odot}$ for Model 1 and 2, respectively. The mass is determined by $\dot{M}_\mathrm{inf}$ and the outer radius of the structure. The outer radius, however, is not well constrained. An $r=12 000\,\mathrm{au}$ model matches the data as well, but the mass of the envelope would then only be $\approx0.14\, \rm M_{\odot}$. Our mass estimates are similar to the derived envelope masses for Class I objects reported by \citet{Kospal2017} in their sample of outbursting sources. In addition, \citet{Furlan2016} derived $T_\mathrm{bol}=492\,\rm K$, which further indicates that V883 Ori is a Class I object. Our simple model is consistent with the Class I evolutionary stage (see also Appendix~\ref{sec:modelcomparison}), but current mass estimates of the envelope and the disk are uncertain and require more detailed modeling of multi-wavelength data to precisely constrain the evolutionary stage of V883 Ori.
\section{Discussion}\label{sec:discussion}

Our models are consistent with a large \mbox{($r\sim10000$ au)} envelope-like structure traced by the lower-velocity feature centered at the systemic-velocity of the CO lines in the APEX data (Fig.~\ref{fig:spec}). We note that the $^{12}$CO APEX data show strong emission at radial distances $\gtrsim10000\,\mathrm{au}$, which may be related to the central envelope or outflow material. A larger envelope model, however, would be inconsistent with the more compact $^{12}$CO(6--5) emission. A model that matches $^{13}$CO would require twice as massive of an envelope, but is inconsistent with the $^{12}$CO emission. This more massive envelope produces a $^{13}$CO line profile that is broader than observed and the line becomes optically thick. 

Considering the high-resolution ALMA data of \citet{ruiz17b} and the APEX observations in context of our model implies that the $^{13}$CO(3--2) line is under-predicted because our model does not include outflows. The ALMA data trace the inner $<10{''}$ region ($\sim5000\,\mathrm{au}$) and show a prominent outflow structure. However, due to limited baseline coverage, the ALMA observations are prone to spatial filtering. For example, the red-shifted outflow visible in $^{13}$CO ALMA observations is not seen in $^{12}$CO. The APEX observations do not resolve this inner region but clearly show large scale envelope structure ($r\gtrsim5000\,\mathrm{au}$), which is likely spatially filtered out in the ALMA data. The $^{13}$CO(3--2) line is optically thin in our model with $\tau_\mathrm{line}\lesssim 1$, but the modeled emission is nearly a factor of five too weak compared to the data. The main isotopologue lines have optical depths $\tau>10$ in our model, consistent with the observations of \citet{ruiz17b} where the inner disk structure becomes visible in $^{13}$CO(2--1) but not in $^{12}$CO(2--1). If the lower-velocity $^{13}$CO(3--2) component in the envelope is indeed optically thin, then the APEX observation of $^{13}$CO(3--2) might be more dominated by outflow emission than the other optically thick lines. This indicates that our model is consistent with the ALMA data but fails to reproduce the $^{13}$CO emission as it traces predominantly outflows not included in our modelling. Regardless, the observed CO emission of V883 Ori is likely a superposition of outflow and envelope emission as the characteristic outflow velocity is only $0.65\,\mathrm{km\,s^{-1}}$ \citep{ruiz17b}, hence most of the outflow emission is likely in the lower-velocity component that we model as a pure envelope structure. However, as the ALMA observations miss the large scale emission and the APEX data do not resolve the outflow region, it is not possible with the currently available data to clearly disentangle and quantify those two contributions.

If V883 Ori's total luminosity ($\rm 200~L_{\odot}$) is indeed dominated by accretion, then the current mass accretion rate onto the star is $\rm \dot{M}_{accr}\sim 1-2 \times 10^{-5}~M_{\odot}~{yr^{-1}}$ \citep[assuming the protostar is $\sim 2.3\,\rm R_{\odot}$ and $\sim 1.3\,\rm M_{\odot}$ as determined by][]{cieza16}. ALMA observations suggest an outflow rate of $\rm \dot{M}_{outf}\sim2\times10^{-6}~M_{\odot}~{yr^{-1}}$ \citep{ruiz17b} and the mass-infall rate onto the envelope from our model is $\rm \dot{M}_{inf}\approx 1-2 \times10^{-6}~M_{\odot}~{yr^{-1}}$. This is consistent with the episodic accretion scenario, where $\rm \dot{M}_{accr}< \dot{M}_{inf}$ in the quiescent stage allowing material to accumulate in the disk and trigger the outburst. Consequently, $\rm \dot{M}_{accr}$ abruptly increases yielding $\rm \dot{M}_{accr}>\dot{M}_{inf}$. The observed $\rm \dot{M}_{outf}/\dot{M}_{accr}$ ratio is consistent with the canonical value of $\sim0.1$ \citep{pelletier}. However, the high value of $\rm \dot{M}_{outf}$ suggests the outflow is a result of an accretion outburst, as otherwise $\rm \dot{M}_{outf}/\dot{M}_{accr}\gtrsim1$, assuming that the quiescent accretion rate is $\rm \lesssim \dot{M}_{inf}$. \citet{ruiz17b} constrained the dynamical timescale of the outflow to $\sim10^{4}\,\mathrm{yr}$. This timescale implies the outflow structure may not be related to the ongoing outburst. We infer that the outflow may instead be a consequence of a previous outburst from $\gtrsim10^{4}\,\mathrm{yr}$ ago with properties similar to the current outburst. Such a scenario is consistent with episodic accretion models \citep[e.g.,][]{vorobyov15} and derived burst frequencies \citep[e.g.,][]{Jorgensen2015b}. This implies the observed combination of envelope, outflow, and diffuse material may be predominately shaped by multiple outbursts.

Conversely, if the outflow/inflow rates are taken at face value, and assuming the quiescent accretion rate is well represented by the mass infall rate in the envelope, we find the ratio to be of order unity, much higher than the canonical value of $\sim0.1$ \citep{pelletier}. This implies that either the mass outflow rate has increased significantly or the mass infall rate has decreased significantly. The discrepancy could, however, arise from a deviation from the assumed morphology or contamination from the surrounding molecular cloud. As can be seen in Fig.\,\ref{fig:int_map}, the $^{12}$CO(6--5) line traces a more compact region and does not show as much large scale emission as the other $^{12}$CO lines. Assuming that $^{12}$CO(6--5) is a better tracer of the envelope implies that the other $^{12}$CO lines may be contaminated by large scale emission from the molecular cloud. This is consistent with Model 1, which better represents the large scale structure, but over-predicts the flux in the central region. 

High spatial and velocity resolution maps are necessary to further constrain the envelope parameters and obtain a single model that reproduces all the molecular species. Single pointings with ALMA \citep[e.g.,][]{cieza16, ruiz17b} will resolve out most of the large scale structure in the envelope, therefore $\gtrapprox30''$ maps with uniform sensitivity are necessary to ensure large scale structure is preserved. Furthermore, the  spatial resolution of the APEX data is not sufficient to see a clear difference in the shapes of the spectral profiles that might indicate the different origin of the $^{12}$CO and $^{13}$CO lines. Having a combination of high spatial and velocity resolution over a large area with ALMA and future facilities such as the ngVLA \citep[e.g.,][]{white18} will allow for the detection of any asymmetries in the envelope as well as accurately disentangle the contribution from inflows, outflows, and disk emission at all velocities.

\section{Summary}\label{sec:summary}

We presented APEX FLASH$^{+}$ and SEPIA B9 science verification observations of the young outbursting star V883 Ori. We mapped and modeled the $^{12}$CO(6--5), $^{12}$CO(4--3), $^{12}$CO(3--2), and $^{13}$CO(3--2) lines of the envelope. Each line was resolved and detected at a high high signal-to-noise. We find the envelope is well represented by a spherical model with a $\gtrsim10000\,\mathrm{au}$ radius, a mass of $\rm M_{env}\approx 0.2-0.4 \,M_{\odot}$, and a mass-infall rate of $\rm \dot{M}_{inf}= 1-2 \times10^{-6} \,M_{\odot}\,{yr^{-1}}$. The model does not, however, reproduce $^{13}$CO(3--2) and we conclude that this line is likely dominated by emission from the outflow structure apparent in high-resolution ALMA data \citep{ruiz17a}. The large-scale structure in the $^{12}$CO lines, that we interpret as an envelope, is likely spatially filtered out in the ALMA observations. The combination of the inferred accretion rate and the observed envelope and outflow structures may be predominately shaped by multiple outbursts consistent with an episodic accretion scenario.

\acknowledgments{

We thank the anonymous referee for feedback that improved the manuscript. We thank A. Hacar for discussions on the interpretation of the data. This publication is based on data acquired under programs 0102.F-9802(A) and 098.F-9505(A) with the Atacama Pathfinder Experiment (APEX). APEX is a collaboration between the Max-Planck-Institut f\"ur Radioastronomie, the European Southern Observatory, and the Onsala Space Observatory. This project received funding from the European Research Council (ERC) under the European Union's Horizon 2020 research and innovation program, grant agreement No 716155 (SACCRED). CH.R. acknowledges funding from the Netherlands Organisation for Scientific Research (NWO) TOP-1 grant project number 614.001.552. CH.R. and E.V. acknowledge funding by the Austrian Science Fund (FWF): project number I2549-N27. M.A. and A.P. acknowledge funding by the Swiss National Science Foundation (SNSF): project number: 200021L\_163172}. T.Cs. acknowledges support from the \emph{Deut\-sche For\-schungs\-ge\-mein\-schaft, DFG\/}  via the SPP (priority program) 1573 ``Physics of the ISM". We thank the Center for Information Technology of the University of Groningen for their support and access to the Peregrine high performance computing cluster. 

\vspace{5mm}
\facilities{APEX(SEPIA, FLASH+)}


\software{GILDAS, P{\small RO}D{\small I}M{\small O} \citep{Woitke2009,Woitke2016,Rab2017}
          }



\begin{appendix}\label{sec:app}

\section{Velocity channel maps}\label{sec:vel_app}
The velocity channels maps for each species are shown in Figs.\,\ref{fig:vel_chan65}, \ref{fig:vel_chan43}, \ref{fig:vel_chan32}, and \ref{fig:vel_chan1332} below. As the $^{12}$CO(6--5) line has the lowest velocity resolution of the line observed with $\rm 0.20~km~s^{-1}$, the other lines are re-binned so that every species has the same velocity resolution. In each figure, $3\sigma$ contours are shown where $\sigma$ is the $\sigma_{RMS}$ of the line-free channels. For each species, we find the emission around $\rm 4.1~km~s^{-1}$ to be more spatially concentrated around the location of V883 Ori, which is indicated by the black plus in the center of each image. This is consistent with emission from an envelope structure as detailed in Sec.\,\ref{sec:env}. The higher-velocity emission peaks seen in Fig.\,\ref{fig:spec} are significantly less spatially concentrated for all species and may be consistent with a combination of diffuse emission from the molecular cloud and outflow material.

\begin{figure*}
    \centering
    \includegraphics[width=\textwidth]{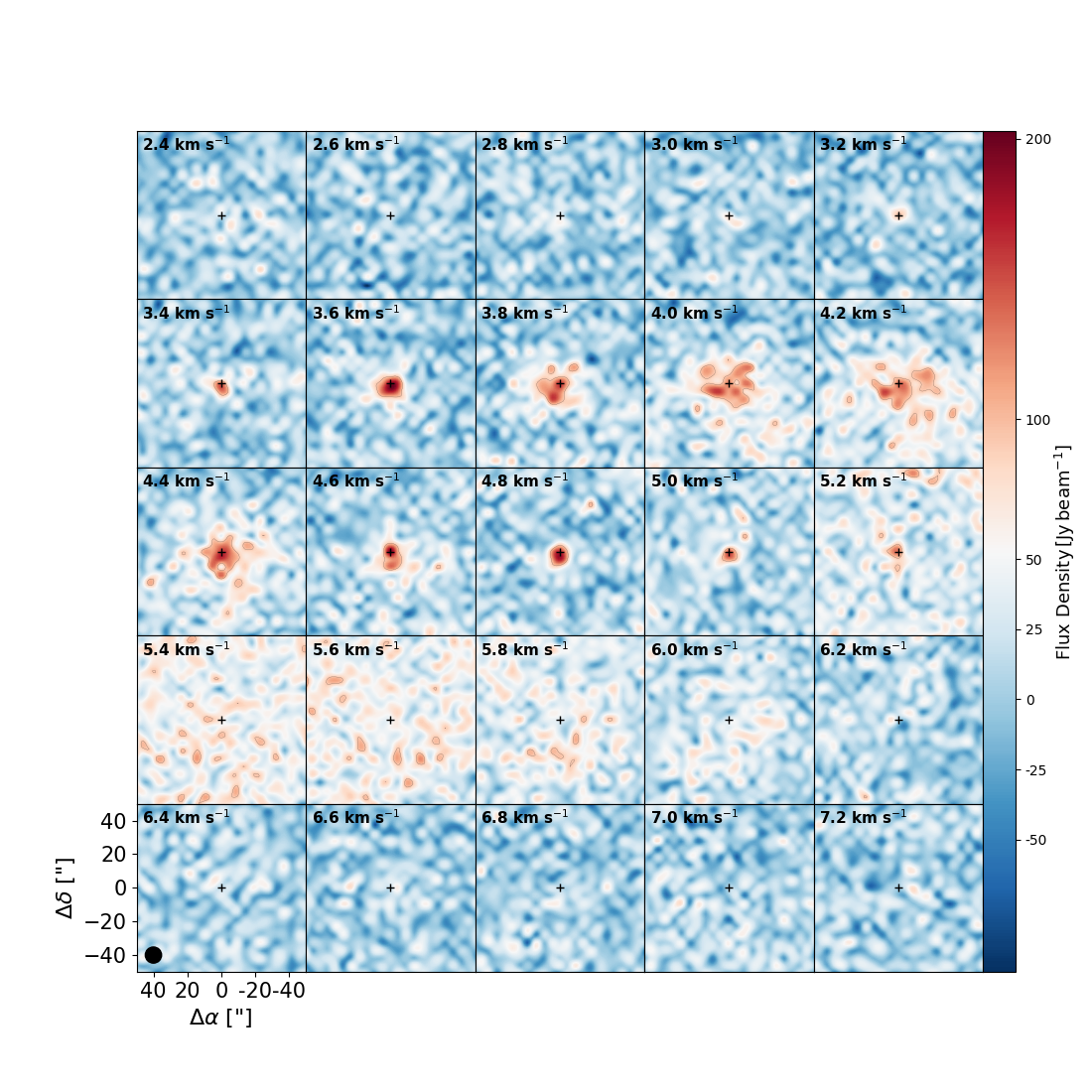}
    \caption{$^{12}$CO(6--5) velocity channel map. The selected channels highlight how the emission from V883 Ori's envelope is predominately located at velocities $\rm <5~km~s^{-1}$ (the lower-velocity peak in all spectra in Fig.\,\ref{fig:spec}) while the emission at velocities $\rm >5~km~s^{-1}$ is likely due to outflow material or diffuse emission from the molecular cloud. The $3\sigma$ contours are given by the black lines with $\sigma = 30 \rm ~ Jy~ beam^{-1}$ ($0.38 \rm ~ K ~ beam^{-1}$). The position of the star is denoted by the black plus and the beam size is denoted by the black circle.}\label{fig:vel_chan65}
    
\end{figure*}

\begin{figure*}
    \centering
    \includegraphics[width=\textwidth]{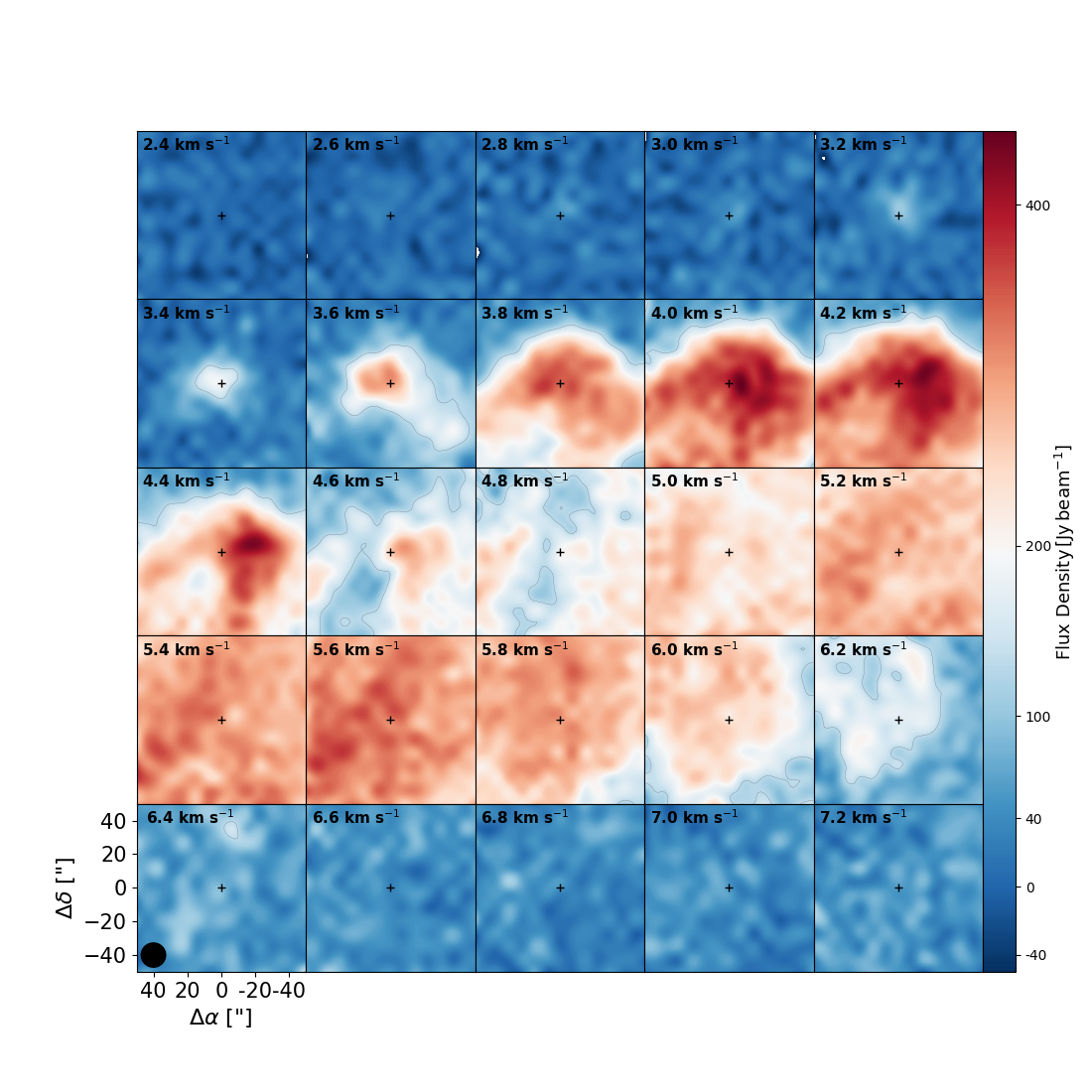}
    \caption{$^{12}$CO(4--3) velocity channel map with channels re-binned to $\rm 0.2~km~s^{-1}$. The selected channels highlight how the emission from V883 Ori's envelope is predominately located at velocities $\rm <5~km~s^{-1}$ (the lower-velocity peak in all spectra in Fig.\,\ref{fig:spec}) while the emission at velocities $\rm >5~km~s^{-1}$ is likely due to outflow material or diffuse emission from the molecular cloud. The $3\sigma$ contours are given by the black lines with $\sigma = 42 \rm ~ Jy~ beam^{-1}$  ($0.88 \rm ~ K ~ beam^{-1}$). The position of the star is denoted by the black plus and the beam size is denoted by the black circle.}\label{fig:vel_chan43}
    
\end{figure*}

\begin{figure*}
    \centering
    \includegraphics[width=\textwidth]{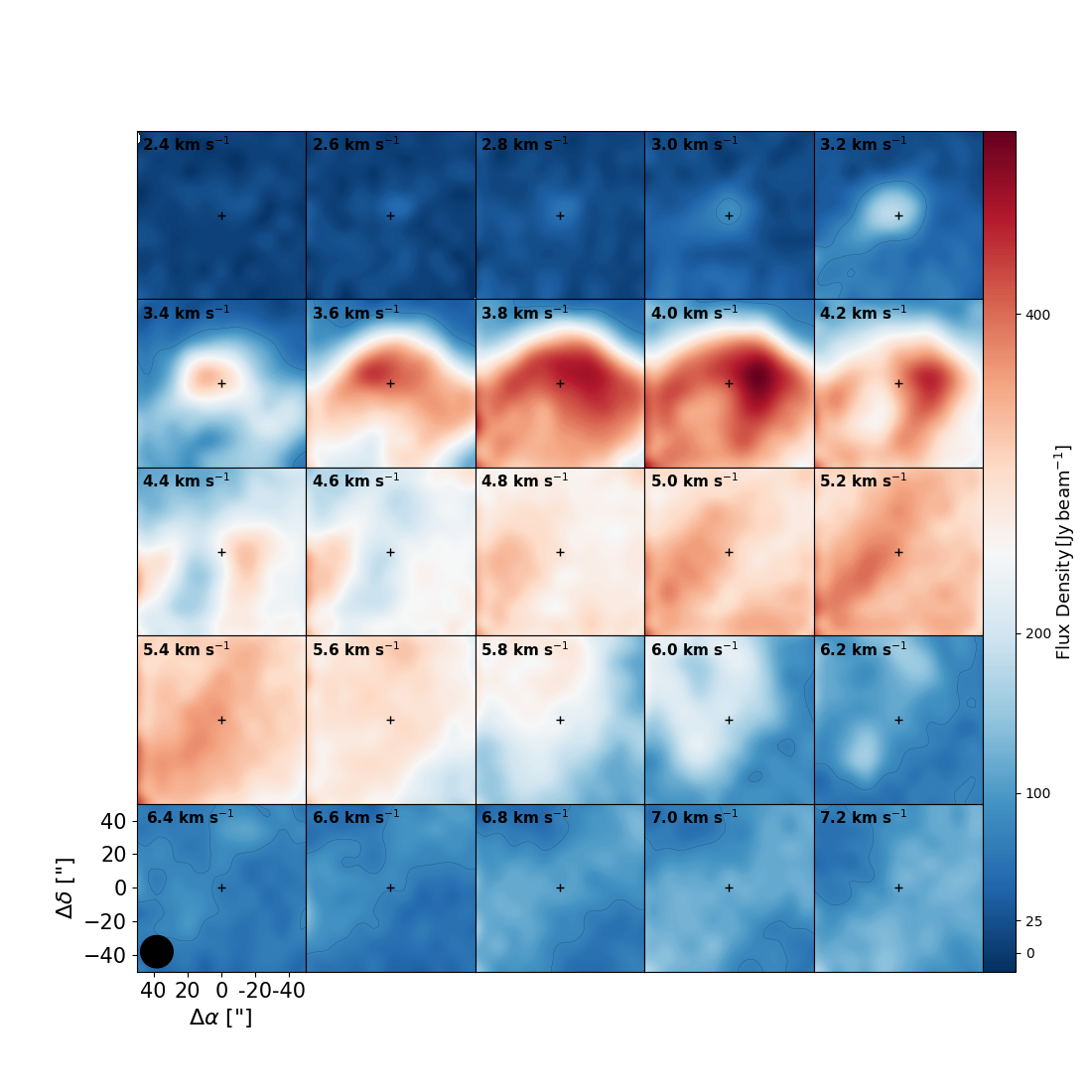}
    \caption{$^{12}$CO(3--2) velocity channel map with channels re-binned to $\rm 0.2~km~s^{-1}$.. The selected channels highlight how the emission from V883 Ori's envelope is predominately located at velocities $\rm <5~km~s^{-1}$ (the lower-velocity peak in all spectra in Fig.\,\ref{fig:spec}) while the emission at velocities $\rm >5~km~s^{-1}$ is likely due to outflow material or diffuse emission from the molecular cloud. The $3\sigma$ contours are given by the black lines with $\sigma = 22 \rm ~ Jy~ beam^{-1}$   ($0.54 \rm ~ K ~ beam^{-1}$). The position of the star is denoted by the black plus and the beam size is denoted by the black circle.}\label{fig:vel_chan32}
    
\end{figure*}

\begin{figure*}
    \centering
    \includegraphics[width=\textwidth]{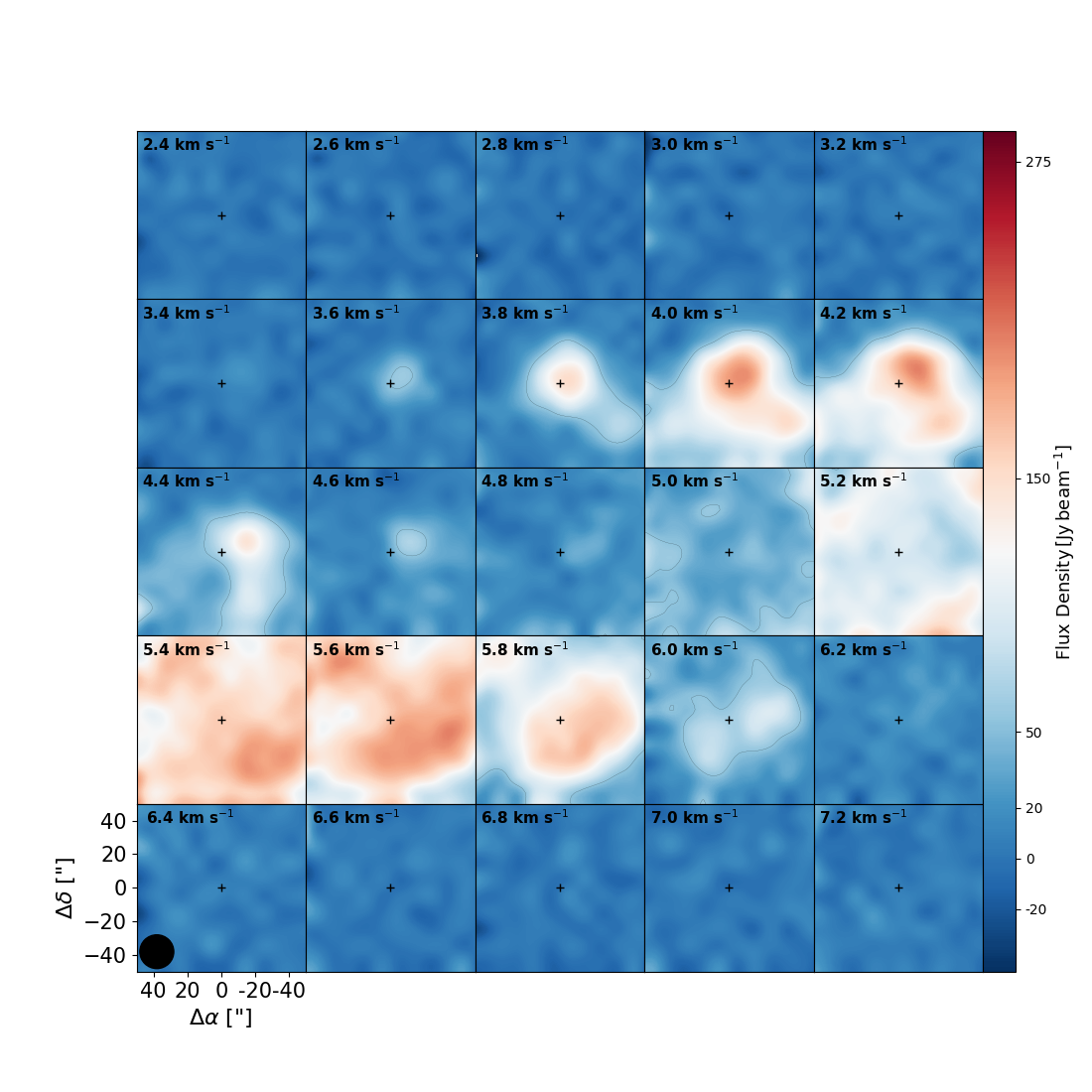}
    \caption{$^{13}$CO(3--2) velocity channel map with channels re-binned to $\rm 0.2~km~s^{-1}$. The selected channels highlight how the emission from V883 Ori's envelope is predominately located at velocities $\rm <5~km~s^{-1}$ (the lower-velocity peak in all spectra in Fig.\,\ref{fig:spec}) while the emission at velocities $\rm >5~km~s^{-1}$ is likely due to outflow material or diffuse emission from the molecular cloud. The $3\sigma$ contours are given by the black lines with $\sigma = 18 \rm ~ Jy~ beam^{-1}$   ($0.44 \rm ~ K ~ beam^{-1}$). The position of the star is denoted by the black plus and the beam size is denoted by the black circle.}\label{fig:vel_chan1332}
\end{figure*}

\section{Modelling procedure and Comparison to Furlan et al. (2016)}
\label{sec:modelcomparison}
In this section we provide further details on our modelling procedure for the APEX data and a comparison to models of the continuum emission of V883 Ori. For the description of the modelling code see Sect~\ref{sec:env}.

For our modeling, we at first computed a small grid of models varying the mass infall rate $\dot{M}_\mathrm{inf}$ in the range of $5\times10^{-7} - 5\times10^{-5}\, \rm M_{\odot} ~ yr^{-1}$ and the central luminosity $L$ in the range of $100-400\, \rm L_{\odot}$. We next sampled parameters around the best matching model of the initial grid varying only the $\dot{M}_\mathrm{inf}$. Our models do not include a disk component as the APEX data is not sensitive to the disk emission (see Sec.~\ref{sec:env}) and to minimize the required computational time. From this exploration of parameter space, we chose the Model 1 and Model 2, as described in Sec.\,\ref{sec:env}.

For the two chosen models, we explored including a disk component with $M_\mathrm{disk}=0.2\,\rm M_{\odot}$ and a radius of $200\,\mathrm{au}$. The impact of the disk component on the synthetic line observables we compare to the APEX data is not significant and does not affect any of our conclusions. However, the disk component is the dominant source of continuum emission at wavelengths $\gtrsim100\,\mathrm{\mu m}$. We note that it is outside the scope of this paper to fit all the available data (including the continuum), and instead aim to provide a reasonable model for the gaseous envelope structure of V883 Ori. Nevertheless, it is useful to compare our modelling results also to pure continuum models.

\citet{Furlan2016} modeled the SED of a large number of protostars in Orion also including V883 Ori (HOPS-376). Their modelling approach also assumes a fixed rotating envelope structure and a disk component. However, they note that they did not vary the disk parameters in their modelling and assumed that the data is not sensitive to the disk. In Table~\ref{tab:modcomparison} we compare $L$, $\dot{M}_\mathrm{inf}$, and the continuum fluxes at $350\,\mathrm{\mu m}$ and $870\,\mathrm{\mu m}$. To derive $\dot{M}_\mathrm{inf}$ from \citet{Furlan2016}, we used their Eq.~2 with $M_\mathrm{*}=1.3\,\rm M_{\odot}$. Our results are in good agreement, however their $\dot{M}_\mathrm{inf}$ is $2-4$ times that of our models. In \citet{Furlan2016}, $\dot{M}_\mathrm{inf}$ is not a direct input parameter but depends on the stellar mass, which was fixed to $M_{*}=0.5\,\rm M_{\odot}$, and results in $\dot{M}_{\rm inf}\approx 2.5\times10^{-6}\,\rm M_{\odot}\,{yr^{-1}}$. From Table~\ref{tab:modcomparison}, it is also apparent that in our model the continuum fluxes are dominated by the disk. \citet{Furlan2016} do not discuss if this is also the case for their V883 Ori model.

\citet{Furlan2016} classified V883 Ori as a flat spectrum source and derived $L_\mathrm{bol}=217.88\,\mathrm{L_\odot}$ and $T_\mathrm{bol}=492\,\mathrm{K}$ from the observational data considered for their modelling. We note that a bolometric luminosity of $400\,\mathrm{L_\mathrm{\odot}}$ \citep{strom93} is commonly adopted for V883 Ori, however it is not apparent how \citet{strom93} derived this value. It might be the case that $L_\mathrm{bol}$ of V883 Ori dropped over the last decade. Regardless, our model is consistent with the values reported by \citet{Furlan2016}.

\begin{table}
\begin{center}
\caption{Comparison of our models to the continuum model and selected photometric fluxes of \citet{Furlan2016}. The parameters are described in the Sec.\,\ref{sec:env} and \ref{sec:modelcomparison} and in \citet{Furlan2016}.}\label{tab:modcomparison}
\bigskip
\begin{tabular}{c|c c c c }
\hline
  Model & $L$ & $\dot{M}_\mathrm{inf}$ & 350 $\mathrm{\mu m}$ Flux & $870\,\mathrm{\mu m}$ Flux  \\
   & $[L_\mathrm{\odot}]$ & $[M_{\odot}\,\mathrm{yr^{-1}}]$ & [Jy] & [Jy] \\
\hline\hline
 \citet{Furlan2016} & 194.67 &  $\approx4\times10^{-6}$  & $11.3\pm4.5$ & $1.2\pm0.2$ \\
 Model 1 & 200 &  $2\times10^{-6}$  & $3.4$ & $0.2$ \\
 Model 1 + disk & 200 &  $2\times10^{-6}$  & $14.3$ & $1.6$ \\
 Model 2 & 200 &  $1\times10^{-6}$  & $1.9$  & $0.1$  \\
 Model 2 + disk & 200 &  $1\times10^{-6}$ & 12.5 & 1.46
 \end{tabular}
\end{center}
\end{table}

\end{appendix}


\begin{thebibliography}{}


\bibitem[Arce \& Sargent(2006)]{arce06}Arce, H.G. \& Sargent, A.I., 2006, ApJ, 646(2), p.1070.

\bibitem[Audard et al.(2014)]{audard14}Audard, M., \'Abrah\'am, P., Dunham, M.M., et al., 2014, Protostars \& Planets VI, 387.

\bibitem[Baryshev et al.(2015)]{baryshev15}Baryshev, A.M., Hesper, R., Mena, F.P., et al., 2015, A\&A, 577, A129

\bibitem[Belitsky et al.(2018)]{belitsky18}Belitsky, V., Lapkin, I., Fredrixon, M., et al., 2018, A\&A, 612, A23

\bibitem[Cassen \& Moosman(1981)]{cassen81}Cassen, P. \& Moosman, A., 1981, Icarus, 48(3), pp.353-376.

\bibitem[Cieza et al.(2016)]{cieza16}Cieza, L.A., Casassus, S., Tobin, J., et al., 2016, Nature, 535(7611), p.258.

\bibitem[Cieza et al.(2018)]{Cieza2018}
{Cieza}, L.A., {Ru{\'\i}z-Rodr{\'\i}guez}, D., {Perez}, S., et al., 2018, \mnras, 474, 4347-4357

\bibitem[Furlan et al.(2016)]{Furlan2016}
{Furlan}, E., {Fischer}, W.~J., {Ali}, B. et al., 2016, ApJS, 224, 5

\bibitem[Green et al.(2006)]{green06}Green, J.D., Hartmann, L., Calvet, N., et al. 2006, ApJ, 648, 1099

\bibitem[Greene et al.(2008)]{greene08}Greene, T. P., Aspin, C., \& Reipurth, B. 2008, AJ, 135,1421

\bibitem[G\"usten et al.(2006)]{gusten06}G{\"u}sten, R., Nyman, L.{\AA}., Schilke, P. et al., 2006, A\&A, 454(2), pp.L13-L16.

\bibitem[Hacar et al.(2018)]{Hacar2018}
{Hacar}, A. and {Tafalla}, M. and {Forbrich}, J. and {Alves}, J. and {Meingast}, S. and {Grossschedl}, J. and {Teixeira}, P.~S. 2018, \aap, 610, A77

\bibitem[Hartmann \& Kenyon(1996)]{hartmann96}Hartmann, L., \& Kenyon, S. J. 1996, ARA\&A, 34, 207

\bibitem[Herbig(1966)]{herbig66}Herbig, G.H., 1966, Vistas in Astronomy, 8, pp.109-125.

\bibitem[Herbig(1989)]{herbig89}Herbig G. H., 1989, In ESO Workshop on Low Mass Star Formation and Pre-Main Sequence Objects (B. Reipurth, ed.), pp. 233-246. ESO, Garching

\bibitem[Hesper et al.(2018)]{hesper18}Hesper, R., Khudchenko, A., Lindemulder, M., et al., 2018, proc. 29th ISSTT, March 2018, nrao.edu/meetings/isstt/papers/2018/2018098103.pdf

\bibitem[J{\o}rgensen et al.(2015)]{Jorgensen2015b} J{\o}rgensen, J.~K., Visser, R., Williams, J.~P., et al., 2015, \aap, 579, A23

\bibitem[Klein et al.(2014)]{klein14}Klein, T., Ciechanowicz, M., Leinz, C., et al., 2014, IEEE Transactions on Terahertz Science and Technology, 4(5), pp.588-596.

\bibitem[K\'osp\'al et al.(2017)]{Kospal2017}K\'osp\'al, \'A., \'Abrah\'am, P., Csengeri, T., et al., 2017, \apj, 836, 226

\bibitem[Kounkel et al.(2018)]{kounkel18}Kounkel, M., Covey, K., Suarez, G., et al., 2018, arXiv preprint arXiv:1805.04649.

\bibitem[Lee et al.(2019)]{lee19}Lee, J.E., Lee, S., Baek, G., et al., 2019, Nature Astronomy, p.1.

\bibitem[Menten et al.(2007)]{menten07}Menten, K.M., Reid, M.J., Forbrich, J., et al., 2007, A\&A, 474(2), pp.515-520.

\bibitem[Millan-Gabet et al.(2006)]{millan06}Millan-Gabet, R., Monnier, J.D., Akeson, R.L., et al., 2006, ApJ, 641(1), p.547.

\bibitem[Pelletier \& Pudritz(1992)]{pelletier}Pelletier, G., \& Pudritz, R.E., 1992, ApJ, 394, pp.117-138.

\bibitem[Pickering(1890)]{pickering1890}Pickering, E.C., 1890, The Observatory, 13, pp.80-81.

\bibitem[Quanz et al.(2007)]{quanz07}Quanz, S.P., Henning, T., Bouwman, J., et al. 2007, ApJ, 668, 359

\bibitem[Quillen et al.(2005)]{quillen05}Quillen, A.C., Thorndike, S.L., Cunningham, A., et al., 2005, ApJ, 632(2), p.941.

\bibitem[Rab et al.(2017)]{Rab2017}Rab, C., Elbakyan, V., Vorobyov, E., et al., 2017, A\&A,  604, A15
  
\bibitem[Ru\'iz-Rodr\'iguez et al.(2017a)]{ruiz17a}Ru\'iz-Rodr\'iguez, D., Cieza, L.A., Williams, J.P., et al., 2017, MNRAS, 466(3), pp.3519-3532.

\bibitem[Ru\'iz-Rodr\'iguez et al.(2017b)]{ruiz17b}Ru\'iz-Rodr\'iguez, D., Cieza, L.A., Williams, J.P., et al., 2017, MNRAS, 468(3), pp.3266-3276.

\bibitem[Strom \& Strom(1993)]{strom93} Strom, K. M., \& Strom, S. E. 1993, ApJ, 412, L63

\bibitem[Terebey et al.(1984)]{terebey84}Terebey, S., Shu, F.H., Cassen, P, 1984, ApJ, 286, pp.529-551.

\bibitem[Thi et~al.(2011)]{Thi2011}Thi, W.-F., Woitke, P., \& Kamp, I. 2011, \mnras, 412, 711

\bibitem[Ulrich(1976)]{Ulrich1976}Ulrich, R.~K. 1976, ApJ, 210, 377

\bibitem[van't Hoff et al.(2018)]{vant18}van't Hoff, M.L., Tobin, J.J., Trapman, L., et al. 2018, ApJL, 864(1), p.L23

\bibitem[Visser et al.(2015)]{Visser2015} {Visser}, R. and {Bergin}, E.~A. and {J{\o}rgensen}, J.~K. 2015, \aap, 577, A102

\bibitem[Vorobyov \& Basu(2015)]{vorobyov15}Vorobyov, E.I., \& Basu, S., 2015, ApJ, 805(2), p.115

\bibitem[White et al.(2018)]{white18}White, J.A., Audard, M., \'Abrah\'am, P., et al., 2018, Science with a Next Generation Very Large Array, ASP Conference Series, (Vol. 517, p. 177), Edited by Eric Murphy
\bibitem[Williams \& Cieza(2011)]{williams11}Williams, J.P. \& Cieza, L.A., 2011, ARA\&A, 49, pp.67-117.

\bibitem[Woitke et al.(2009)]{Woitke2009}Woitke, P., Kamp, I., \& Thi, W.-F., 2009, A\&A, 501, 383

\bibitem[Woitke et~al.(2011)]{Woitke2011}Woitke, P., Riaz, B., Duch{\^e}ne, G., et al., 2011, \aap, 534, A44

\bibitem[Woitke et~al.(2016)]{Woitke2016}Woitke, P., Min, M., Pinte, C., et al., 2016, A\&A, 586, A103



\end{thebibliography}
\end{document}